\documentclass[letter,twocolumn]{jpsj3} 
\usepackage{ulem}
\usepackage{color}

\title{
Optical Evidence of Itinerant-Localized Crossover of $4f$ Electrons in Cerium Compounds
}
\author{
Shin-ichi \textsc{Kimura}$^1$\thanks{E-mail address: kimura@fbs.osaka-u.ac.jp}, 
Yong Seung \textsc{Kwon}$^2$,
Yuji \textsc{Matsumoto}$^3$,
Haruyoshi \textsc{Aoki}$^4$,\\
and 
Osamu \textsc{Sakai}$^5$
}
\inst{
$^1$FBS and Department of Physics, Osaka University, Suita, Osaka 565-0871, Japan\\
$^2$Department of Emerging Materials Science, DGIST, Daegu 711-873, Republic of Korea\\
$^3$Department of Engineering Physics, Nagoya Institute of Technology, Nagoya 466-8555, Japan\\
$^4$Department of Physics, Tohoku University, Sendai 980-8578, Japan\\
$^5$National Institute for Materials Science, Tsukuba 305-0047, Japan
}
\recdate{
\today
}
\abst{
Cerium (Ce)-based heavy-fermion materials have a characteristic double-peak structure (mid-IR peak) 
in the optical conductivity [$\sigma(\omega)$] spectra 
originating from the strong conduction ($c$)--$f$ electron hybridization.
To clarify the behavior of the mid-IR peak at a low $c$-$f$ hybridization strength, 
we compared the $\sigma(\omega)$ spectra of the isostructural antiferromagnetic and heavy-fermion Ce compounds
with the calculated unoccupied density of states and the spectra obtained from the impurity Anderson model.
With decreasing $c$-$f$ hybridization intensity, 
the mid-IR peak shifts to the low-energy side owing to the renormalization of the unoccupied $4f$ state,
but suddenly shifts to the high-energy side owing to the $f$-$f$ on-site Coulomb interaction 
at a slight localized side from the quantum critical point (QCP).
This finding gives us information on the change in the electronic structure across QCP.
}
\begin{document}
\maketitle
%
%
Rare-earth intermetallic compounds, especially cerium (Ce) and ytterbium (Yb) compounds, 
have a variety of physical properties of nonmagnetic heavy fermion (HF) 
owing to the Kondo effect with large effective mass and magnetic ordering (MO) 
caused by the Ruderman-Kittel-Kasuya-Yosida (RKKY) interaction.~\cite{Hewson1993}
The change between the nonmagnetic HF and MO states originates from the intensity of the hybridization 
between conduction and $4f$ electrons, namely, $c$-$f$ hybridization.
The phase diagram is known as the Doniach phase diagram.~\cite{Doniach1977}
At the border between the HF and MO states, there is a phase transition at zero temperature, 
namely, the quantum critical point (QCP).
So far, near the QCP, interesting physical properties such as those of non-BCS superconductors and non-Fermi liquids 
owing to charge and magnetic fluctuations have been observed and investigated thoroughly.

\begin{figure}[t]
\begin{center}
\includegraphics[width=0.35\textwidth]{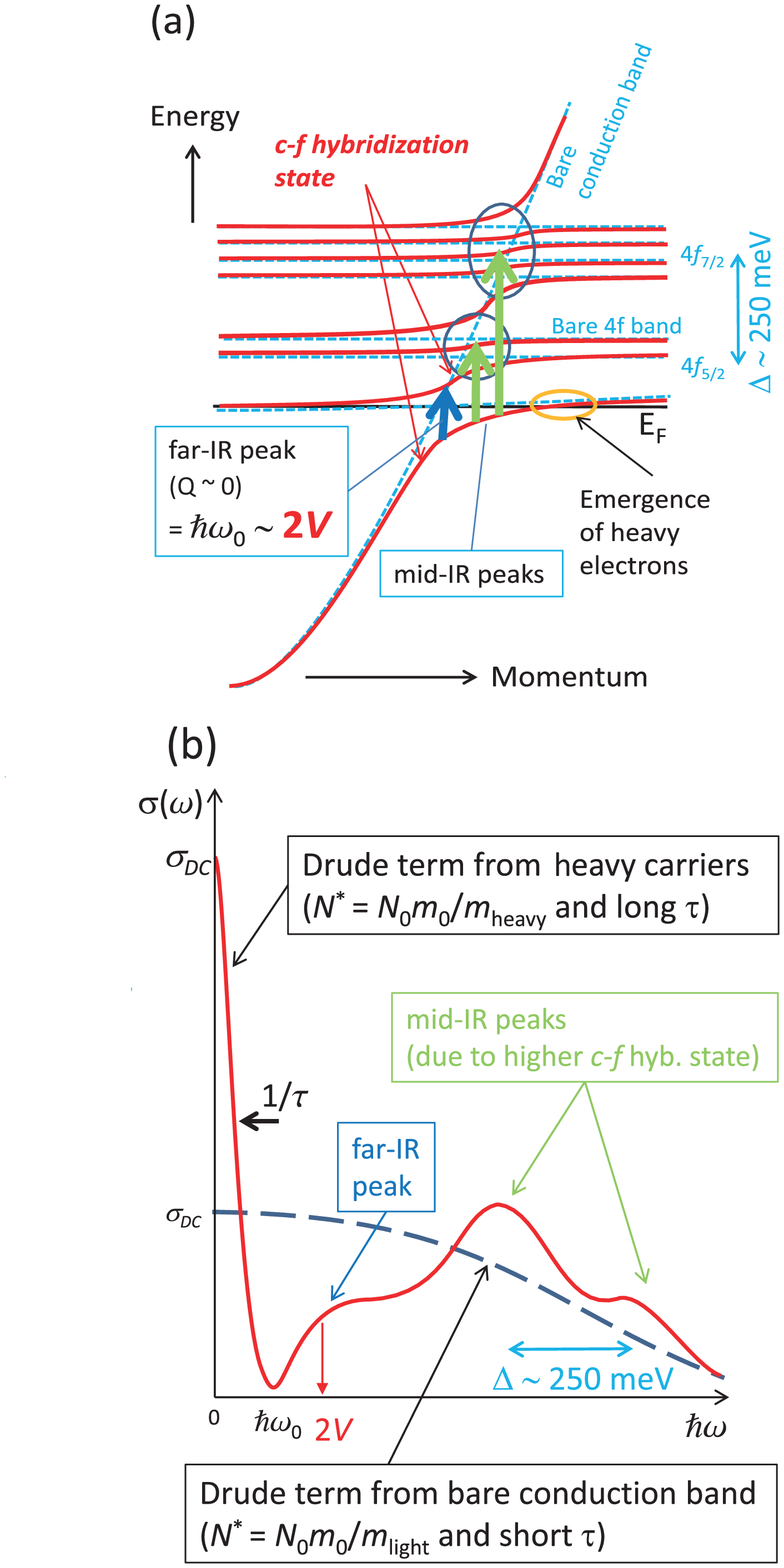}
\end{center}
\caption{
(Color online)
(a) Schematic of electronic structure of Ce compounds with $c$-$f$ hybridization.
Far-IR and mid-IR peaks in optical conductivity [$\sigma(\omega)$] spectra of Ce compounds are expected from the electronic excitation indicated by arrows.
(b) Schematic $\sigma(\omega)$ spectra at high and low temperatures depicted by dashed and solid lines, respectively.
}
\label{fig:schematic}
\end{figure}
The electronic structure as well as the band dispersion of the $c$-$f$ hybridization state, 
which is depicted in Fig.~\ref{fig:schematic}a, can be obtained directly by momentum-resolved experiments such as angle-resolved photoemission (ARPES)~\cite{Im2008} and inelastic neutron scattering (INS).~\cite{Christianson2006}
Optical conductivity [$\sigma(\omega)$] spectra also provide information on the $c$-$f$ hybridization intensity ($V$).~\cite{Degiorgi1999,Kimura2013}
There are three significant structures in the $\sigma(\omega)$ spectrum at lower temperatures than the Kondo temperature ($T_{\rm K}$) as shown in Fig.~\ref{fig:schematic}b.
The first structure is a very narrow Drude peak at $\hbar\omega=$~0~eV with a heavy mass ($m_{heavy}$) and long relaxation time ($\tau$) owing to coherent heavy carriers, 
the second is a shoulder structure in the far-infrared region (hereafter, far-IR peak), 
and the third is a double-peak structure in Ce compounds with an energy separation of 250~meV, which originates from the spin-orbit splitting of the Ce~$4f$ unoccupied state, in the middle-infrared region (hereafter, mid-IR peak).~\cite{Yb}
The far-IR peak, which is observed at $\hbar\omega\leq$~50~meV, originates from the direct interband transition between the bonding and antibonding states of the $c$-$f$ hybridization state with a gap size of $2V$.~\cite{Iizuka2010,Kimura2011}
The gap has a size of $k_{\rm B}T_{\rm K}$, where $k_{\rm B}$ is the Boltzmann constant, and appears below $T_{\rm K}$.
On the other hand, the mid-IR peak is located at $\hbar\omega\sim$ 100~meV, which is one order of magnitude higher than $V$ and appears below a temperature several times higher than $T_{\rm K}$.
However, it has a universal scaling rule with respect to $V$, so it is related to the $c$-$f$ hybridization.~\cite{Okamura2007}
At a much higher temperature than $T_{\rm K}$, the mid-IR peak disappears because $4f$ states do not hybridize with conduction bands.~\cite{Kimura2006}
This implies that the mid-IR peak originates from the optical transition from the bonding state of the $c$-$f$ hybridization at the Fermi level ($E_{\rm F}$) not to the bare $4f$ states but to the hybridization bands between highly localized $4f$ states and conduction bands as shown in Fig.~\ref{fig:schematic}a.

In materials with sufficiently strong $c$-$f$ hybridization intensity, for instance, YbAl$_2$, the electronic structure can be described by an LDA band calculation because of the weak electron correlation.~\cite{Matsunami2013}
In normal HF materials, the mid-IR peak can be explained by a local-density approximation (LDA) band calculation with a self-energy correction.~\cite{Kimura2009-1,Kimura2009-2}
The renormalization factor $z$ decreases from unity with decreasing $V$ indicating strong renormalization near the QCP.
The mid-IR peak is considered to disappear at $V=0$ because $4f$ electrons are perfectly localized and the strong on-site Coulomb interaction is dominant.
Therefore, the $V$-dependent mid-IR peak behavior is important for the investigation of the effect of the $c$-$f$ hybridization.

In this Letter, to investigate the change in the electronic structure with the $c$-$f$ hybridization from the itinerant HF state to the localized MO state across the QCP, we report the mid-IR peaks of the $\sigma(\omega)$ spectra of the Ce-based ternary intermetallic compounds Ce$M_2$Ge$_2$ ($M =$ Ag, Rh, Ru, Cu, Ni) and CeCu$_2$Si$_2$ with the tetragonal ThCr$_2$Si$_2$-type crystal structure and Ce$M$Ge$_2$ ($M =$ Co, Ni) with the orthorhombic CeNiSi$_2$-type structure as a function of hybridization intensity and compare the results with LDA band calculations and the impurity Anderson model.
Ce$M_2$Ge$_2$s are isostructural materials that are located in the MO ground state ($M =$ Ag, Rh, Ru, Cu) and in the HF state (CeNi$_2$Ge$_2$, CeCu$_2$Si$_2$).~\cite{Endstra1993}
On the other hand, Ce$M$Ge$_2$ also has two states, MO ($M =$ Ni) and HF ($M =$ Co).~\cite{Lee2005}
As a result, in the HF state, the mid-IR peak has been observed but the peak energy is located at a lower energy than that predicted by the band calculations.
The mid-IR peak has also been observed even in CeCu$_2$Ge$_2$ and CeNiGe$_2$, which are located slightly in the MO ground state close to the QCP.
In Ce$M_2$Ge$_2$ ($M =$ Ag, Rh, Ru), which are located at a more localized region than CeCu$_2$Ge$_2$, the mid-IR peak disappears and, instead, a broad peak originating from the $f$-$f$ on-site Coulomb interaction appears and shifts to the high-energy side with decreasing $V$.
The behavior of the mid-IR peak with different $c$-$f$ hybridization strength can be reproduced using the impurity Anderson model.

%
Poly- and single-crystalline samples of Ce$M_2$Ge$_2$ ($M =$ Ag, Rh, Ru, Cu, Ni) were synthesized by a tetra-arc melting method and the surfaces were well polished using 0.3-$\mu$m-grain-size Al$_{2}$O$_{3}$ lapping film sheets for optical reflectivity [$R(\omega)$] measurements.
Near-normal incident $R(\omega)$ spectra of polycrystalline samples were accumulated in a very wide photon-energy range of 2~meV -- 30~eV to ensure accurate Kramers-Kronig analysis (KKA)~\cite{Kimura2013}.
To evaluate the anisotropy of the optical spectra along different crystal axes, we performed a polarized $R(\omega)$ measurement of single-crystalline CeRu$_2$Ge$_2$~\cite{Matsumoto2011} along the $a$- and $c$-axes using linearly polarized light, but no significant difference was observed.
Thus we used the optical spectra of polycrystalline samples for the following analysis and discussion.
To obtain $\sigma(\omega)$ via the KKA of $R(\omega)$, the spectra were extrapolated to below 2~meV with a Hagen-Rubens function and to above 30~eV with the free-electron approximation $R(\omega) \propto \omega^{-4}$~\cite{DG}.
$\sigma(\omega)$ spectra of polycrystalline Ce$M_2$Ge$_2$ as well as single-crystalline CeCu$_2$Si$_2$~\cite{Sichelschmidt2013} and polycrystalline Ce$M$Ge$_2$s ($M =$ Co, Ni)~\cite{Kwon2006} were compared with the unoccupied density of states (DOS) obtained from the LDA band structure calculation including spin-orbit coupling using the {\sc Wien2k} code~\cite{Wien2k}.
All experimental spectra were taken at a temperature of 8--10~K.


\begin{figure}[t]
\begin{center}
\includegraphics[width=0.40\textwidth]{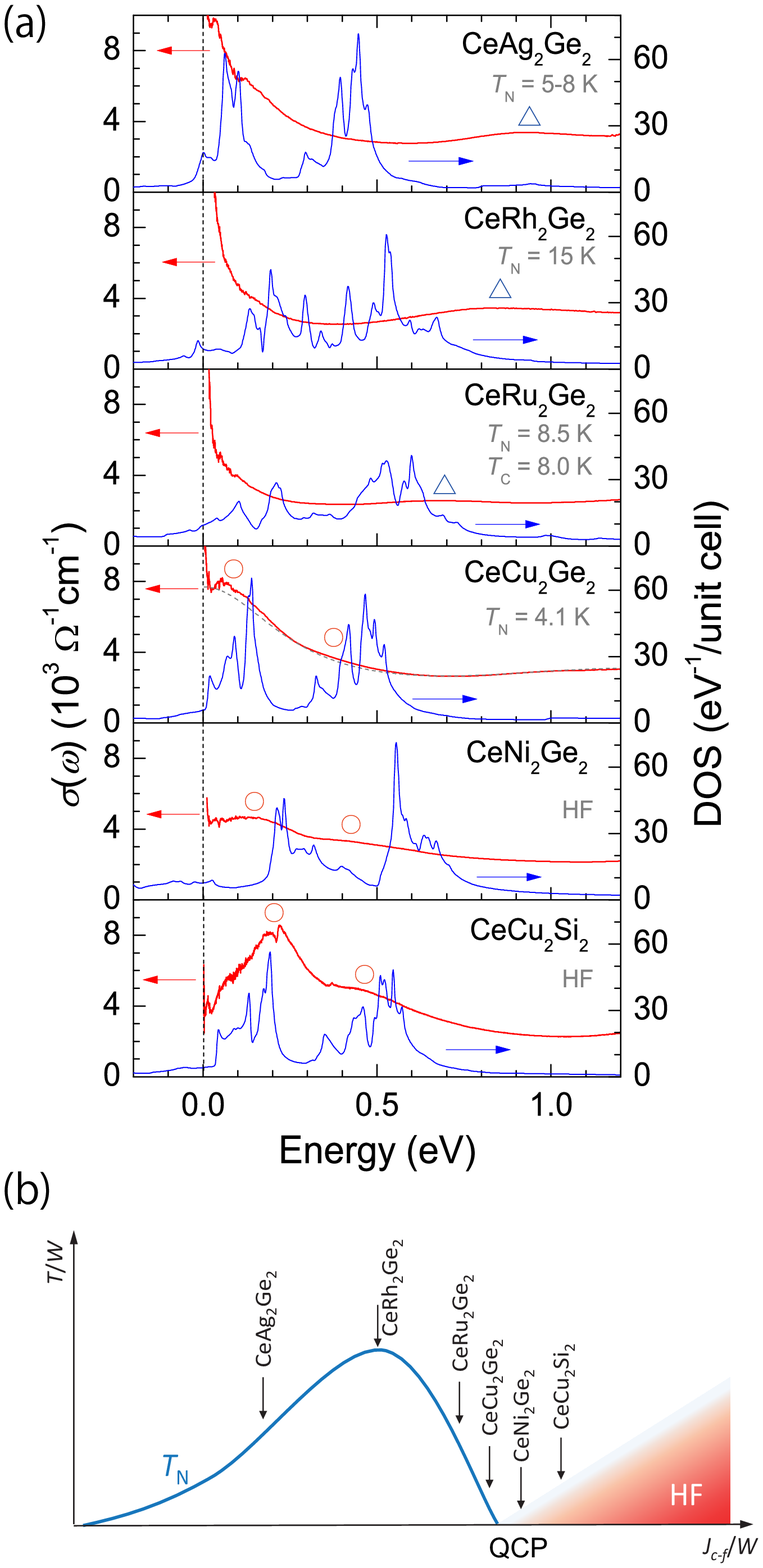}
\end{center}
\caption{
(Color online)
(a) Optical conductivity [$\sigma(\omega)$] spectra of Ce$M_2$Ge$_2$ ($M =$Ag, Rh, Ru, Cu, Ni) and CeCu$_2$Si$_2$ in comparison with the density of states (DOS) derived from an LDA band calculation.
The spectrum of CeCu$_2$Si$_2$ was taken from Ref.~\citen{Sichelschmidt2013}.
The dashed line in CeCu$_2$Ge$_2$ is the background obtained by combining Drude and Lorentz functions as a guide to the eye.
Open circles and open triangles indicate the mid-IR peaks and the optical transitions to the unoccupied $4f$ states with the on-site Coulomb interaction, respectively.
(b) Positions of materials listed in (a) in the Doniach phase diagram derived from Ref.~\citen{Endstra1993}.
}
\label{fig:OC}
\end{figure}
Figure~\ref{fig:OC}a shows the $\sigma(\omega)$ spectra derived from the KKA of $R(\omega)$ spectra and the unoccupied DOS.
A mid-IR double-peak structure (marked by open circles) with an energy separation of $\sim$250~meV is observed in the $\sigma(\omega)$ spectra of CeCu$_2$Si$_2$, CeNi$_2$Ge$_2$, and CeCu$_2$Ge$_2$.
Since the splitting energy of 250~meV is equal to the spin-orbit splitting of Ce~$4f$ states and the peak disappears at very high temperatures (not shown here), the mid-IR peak corresponds to the excitation to the high-energy $c$-$f$ hybridization states as shown in Fig.~\ref{fig:schematic}.
That is, the appearance of the mid-IR peak is evidence of the strong $c$-$f$ hybridization intensity.
In these materials, CeCu$_2$Ge$_2$, which is slightly located in the MO ground state as shown in Fig.~\ref{fig:OC}b, also has a mid-IR peak.
This result suggests that the $c$-$f$ hybridization is still occurring on the slightly localized side from QCP, {\it i.e.}, the HF-like electronic structure with the $c$-$f$ hybridization band is realized in materials of the MO ground state close to the QCP.
This is consistent with the itinerant quantum criticality picture based on the spin fluctuation theory.~\cite{Gegenwart2008}

The mid-IR peak shifts to the low-energy side with decreasing $c$-$f$ hybridization intensity.
In the framework of the photoabsorption process within a one-electron approximation, the lowest absorption peak originates from the excitation from the state at $E_{\rm F}$ to the lowest unoccupied DOS peak.
However, the observed peaks are located at a lower energy than the unoccupied $4f$ states calculated as shown in Fig.~\ref{fig:OC}a.
This implies that the energy position of the unoccupied $4f$ peak is renormalized owing to a self-energy effect.
The renormalization factor becomes small with decreasing $V$.
This is consistent with the enhancement of the effective carrier mass near the QCP.

In the more localized materials of CeRu$_2$Ge$_2$, CeRh$_2$Ge$_2$, and CeAg$_2$Ge$_2$, the mid-IR peak disappears.
This is considered to originate from the weak $c$-$f$ hybridization intensity indicating an electronic structure with a small Fermi surface.~\cite{Gegenwart2008}
However, a broad peak marked by an open triangle appears and shifts to the high-energy side (blue shift) from $M =$~Ru to Ag as shown in Fig.~\ref{fig:OC}a.
The peak cannot be explained by the unoccupied $4f$ DOS obtained from the band calculation.
Other effects, for instance, an on-site Coulomb interaction, should be adopted.

\begin{figure}[t]
\begin{center}
\includegraphics[width=0.35\textwidth]{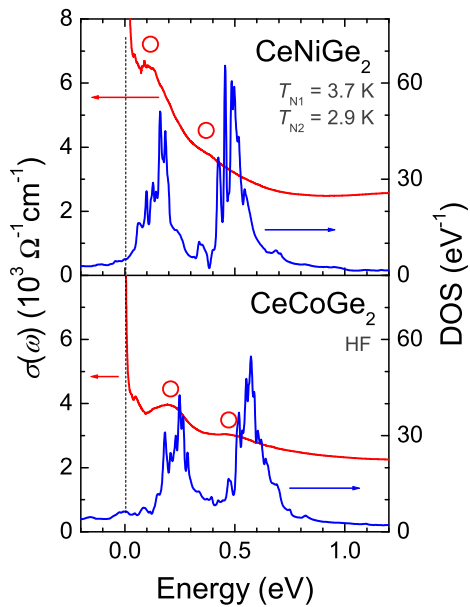}
\end{center}
\caption{
(Color online)
Optical conductivity [$\sigma(\omega)$] spectra of Ce$M$Ge$_2$ ($M =$Ni, Co) in comparison with the DOS derived from LDA band calculations.
The $\sigma(\omega)$ spectra were taken from Ref.~\citen{Kwon2006}.
}
\label{fig:112OC}
\end{figure}
To investigate whether the behavior of the mid-IR peak is observed in other materials, the $\sigma(\omega)$ spectra of Ce$M$Ge$_2$ were compared with the unoccupied DOS as shown in Fig.~\ref{fig:112OC}.
The mid-IR peak energies of CeCoGe$_2$ are consistent with the unoccupied $4f$ peaks because of its heavy-fermion character.
On the other hand, we could recognize that the mid-IR peak in CeNiGe$_2$ is located at a lower energy than the peaks of the unoccupied DOS even though the materials are located in the slightly MO ground state near the QCP.
This result is similar to that of CeCu$_2$Ge$_2$, which is also located in the MO ground state near the QCP.
Therefore, this result supports the fact that the $c$-$f$ hybridization occurs in the slightly localized region from the QCP.

The spectral feature can be explained by the final-state effects.
The ground state is written as $|f^1d^nc^m\rangle$, where the Ce$^{3+}$-$4f$ state ($f$), $M$-$d$ valence band ($d$), and Ce-$5d$ conduction band ($c$) have one, $n$, and $m$ electrons, respectively.
In the case of photoexcitation from the $M$-$d$ state to the Ce-$4f$ state, the optical absorption process can be written as
\[
|f^1d^nc^m\rangle + h\nu \rightarrow |f^2d^{n-1}c^m\rangle.
\]
In the case of a low $c$-$f$ hybridization intensity, the $|f^2d^{n-1}c^m\rangle$ state is stable.
The energy of the final $|f^2d^{n-1}c^m\rangle$ state becomes the sum of the $4f^2$ energy ($\varepsilon_f$), the Coulomb repulsion energy between two $4f$ electrons ($U_{ff}$), and the Coulomb attractive force energy between the $4f$ electron and the $M$-$d$ hole ($-U_{fd}$), namely, $\varepsilon_f+U_{ff}-U_{fd}$.
The total Coulomb energy ($U_{ff}-U_{fd}$) is expected to become large with decreasing $c$-$f$ hybridizaton strength.
Our observation of the blue shift of the broad peak from $M =$~Ru to Ag is consistent with the expectation.

On the other hand, in the case of a strong $c$-$f$ hybridization intensity (HF state), the final state of $|f^2d^{n-1}c^m\rangle$ is not stable because the excited $4f$ electrons can easily move to the $M$-$d$ state and conduction band.
Namely, the photoabsorption final state can be written as
\[
|f^2d^{n-1}c^m\rangle \rightarrow |f^1d^nc^m\rangle + |f^1d^{n-1}c^{m+1}\rangle.
\]
The former cannot be detected because it is the same as the ground state.
The latter is the charge transfer excitation from the $M$-$d$ valence band to the the $c$-$f$ hybridization band, 
which is the same as the optical absorption derived from the LDA band calculation.
Therefore, the $\sigma(\omega)$ spectrum of the materials with strong $c$-$f$ hybridization is considered to be reproduced by the LDA band calculation.~\cite{Kimura2009-1,Kimura2009-2}

\begin{figure}[t]
\begin{center}
\includegraphics[width=0.40\textwidth]{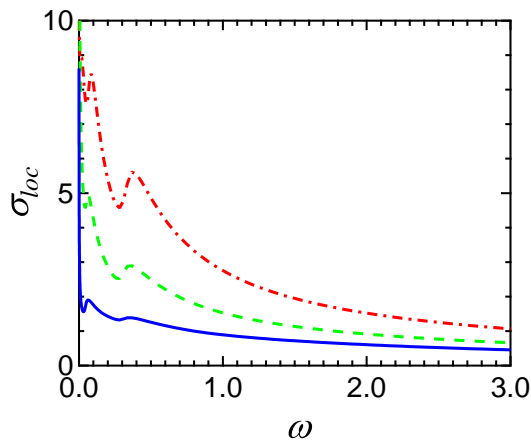}
\end{center}
\caption{
(Color online)
Local optical conductivity ($\sigma_{loc}$) spectra calculated using the single-particle excitation spectra (SPE) of the impurity Anderson model.
The energy levels of the $f$ state are assumed as follows: the energy splitting between the $j=7/2$ multiplet and the ground multiplet $j=5/2$ is 0.3; 
the $j=5/2$ multiplet splits into a ground doublet and an excited quartet with an energy separation of 0.05; the ground $f$ energy level is located at -1.4 
and the $f^{2}$ configuration is removed ({\it i.e.}, infinite $f$-$f$ Coulomb interaction model).
The NCA method is used to calculate SPE.
The hybridization strengths $\pi V^{2}/2D$ are $2.2 \times 10^{-2}$ for the solid line, $2.5 \times 10^{-2}$ for the dashed line, 
and $2.8 \times 10^{-2}$ for the dot-dashed line.
The peaks of the magnetic excitation spectra appear, respectively, at $0.11 \times 10^{-4}$, $0.10 \times 10^{-2}$, and $0.19 \times 10^{-1}$.
These correspond to 0.1, 11, and 210~K, respectively, if we consider a unit to be 1~eV.
The quantity $1/2D$ is the DOS of the conduction band extending from $-D$ to $D$ with width $D=6$.
For the definition of $\sigma_{loc}$, see the text and Ref.~\citen{Takayama}.  
}
\label{fig:NCA}
\end{figure}

To see the correlation effects of the $c$-$f$ hybridization system in detail, 
we show the local optical conductivity ($\sigma_{loc}$) of the impurity Anderson model based on the non-crossing approximation (NCA) calculation in Fig.~\ref{fig:NCA}.
We suppose that $\sigma_{loc}$ is given by a convolution integral of the optical excitation type~\cite{Takayama} between the single-particle excitation spectra (SPE) 
of the $f$ state and the conduction band that has the optical transition matrix with the $f$ state but does not have the hybridization matrix.
We assume that this component of the conduction band has a constant DOS of 1.
The energy splitting between the $j=7/2$ multiplet and the ground multiplet $j=5/2$ of $0.3$ and also the crystal field splitting in the $j=5/2$ multiplet of $0.05$ are introduced.
$\sigma_{loc}$ is large in the strong-hybridization case shown by the dot-dashed line compared with that in the weak-hybridization case shown by the solid line.
The total intensity of the electron excitation component of the SPE ($I_{\rm IPES}$), which corresponds to the total intensity of the inverse photoemission spectra (IPES), is 2.598 for the dot-dashed line case and 0.577 for the solid line case. 
We have the relation $I_{\rm IPES}=14(1-n_{f})$ in the NCA calculation~\cite{NCA}, where $n_{f}$ is the occupation number of $f$ electrons.
The total intensity of the hole excitation component of the SPE, $I_{\rm PES}$, which corresponds to the total intensity of the photoemission spectra (PES), is given as $n_{f}$. 
When the hybridization increases, $I_{\rm IPES}$ increases considerably, while $I_{\rm PES}$ decreases gradually.
The increase in $\sigma_{loc}$ is due to the increase in the transition from the occupied conduction band to the IPES part of the sharp peaks of the $f$ SPE just above the Fermi energy: 
the Kondo resonance peak and the side peaks of the crystal field splitting and spin-orbit splitting.
The peaks of the SPE in the impurity model are interpreted as the peaks of the correlated narrow $f$ band in the lattice system.
Note that the side peaks become conspicuous and show a shift to the high-energy side with increasing hybridization strength.
In the weak-hybridization case (solid line), a flat structure appears around an energy of $\sim 2$. 
This corresponds to the broad peak in the PES owing to the $f^{1} \rightarrow f^{0}$ excitation. 
The Kondo temperature in the manifold of the lowest doublet is estimated to be about 10~K for the case shown by the dashed line.
A low-energy increase in $\sigma_{loc}$ appears in this case, and $\sigma_{loc}$ rapidly becomes weak around this hybridization strength.
The qualitative features seem to accord with the experimental results.
The splitting of the $f$ band due to the MO and/or anomalies due to the QCP will accelerate the change in the shape of the $\sigma(\omega)$ spectrum.

In the present impurity calculation, we tentatively ascribe the structure around 0.1~eV observed in the experiments to the ``crystal field splitting'' of the $f$ state.
In the lattice system, the $f$ electron bands have many peaks in the DOS as shown in Fig.~\ref{fig:OC} obtained from the LDA calculation.
It is probable that the $c$-$f$ hybridization effect in the band causes some side peaks in the experimental $\sigma(\omega)$ spectra.

The Coulomb interaction between an optically excited electron and a hole has not been considered in the present discussion.
This causes the exciton effect in the optical transition~\cite{Takayama}.
The normalized spectral shape of $\sigma(\omega)$ in the low-energy region itself is not changed markedly by this effect, but the absolute intensity is strongly enhanced.
The broad peak due to the $f^{1} \rightarrow f^{0}$ excitation is shifted to the low-excitation-energy side and enhanced from that estimated by the convolution of the SPE.
The origin of the rapid change in the low-energy spectra in the material dependence may be partly attributed to the exciton effect.

These obtained results suggest that the mid-IR peak is a good probe for the $c$-$f$ hybridization intensity.
In particular, the spectral shape of the mid-IR peak markedly changes between CeCu$_2$Ge$_2$ and CeRu$_2$Ge$_2$, which are located slightly in the MO ground state from the QCP.
This implies that the change in the electronic structure from a large Fermi surface caused by strong $c$-$f$ hybridization to a small Fermi surface caused by weak hybridization does not occur at the QCP but in the MO state.
This result is consistent with that the $c$-$f$ hybridization gap opens even in the MO state in CeIn$_3$ studied with $\sigma(\omega)$ spectra under a high pressure~\cite{Iizuka2012} and can be explained by the itinerant quantum criticality picture.~\cite{Gegenwart2008}

%
To summarize, 
the optical conductivity spectra of Ce$M_2$Ge$_2$ ($M =$ Ni, Cu, Ru, Rh, Ag), CeCu$_2$Si$_2$, and Ce$M$Ge$_2$ ($M =$ Ni, Co) have been compared with the corresponding unoccupied DOS to investigate the relationship of the appearance of the mid-IR peak to the $c$-$f$ hybridization intensity.
In the heavy-fermion regime, the mid-IR peak can be explained by LDA calculations with a self-energy correction.
In the localized regime, on the other hand, the mid-IR peak disappears and a broad peak appears at a higher photon energy, 
which cannot be explained by LDA calculations because of the strong $f$-$f$ on-site Coulomb interaction.
The change in the spectral feature occurs at a slightly localized region from the QCP in the Doniach phase diagram.
This is consistent with the itinerant quantum criticality picture based on the spin fluctuation model.
The theoretical study using the impurity Anderson model also supports the change in the spectral feature.

\section*{Acknowledgments}
We would like to thank UVSOR staff members for their technical support.
Part of this work was supported by the Use-of-UVSOR Facility Program of the Institute for Molecular Science.
The work was partly supported by a JSPS Grant-in-Aid for Scientific Research (B) (Grant No. 15H03676).
Y.~S.~K. was supported by the Basic Science Research Program of the NRF (2013R1A1A2009778)
and the Leading Foreign Research Institute Recruitment Program (Grant No. 2012K1A4A3053565).


%

\begin{thebibliography}{99}
\bibitem{Yb}
In Yb compounds, since the spin-orbit splitting energy of the Yb~$4f$ state is about 1.4~eV, one peak originating from Yb-$4f_{7/2}$ is visible in the mid-IR region ($\hbar\omega \sim$~0.1~eV).
The other peak from Yb-$4f_{5/2}$ that should appear at $\hbar\omega\sim$~1.5~eV cannot usually be observed because of the weak hybridization between the Yb-$4f_{5/2}$ state and conduction bands.
\end{thebibliography}
\end{document}